\documentstyle[latexsym,amssymb,epsfig,aps,floats,eqsecnum,amsfonts]{revtex}

\newcommand{\beq}{\begin{equation}}
\newcommand{\eeq}{\end{equation}}

\begin{document}

\title{Gravitational radiation from cosmic (super)strings: bursts, stochastic background, and observational windows}
\author{Thibault Damour$^{1)}$ and Alexander Vilenkin$^{2)}$}
\address{$^{1)}$ {\sl Institut des Hautes Etudes Scientifiques, 91440 Bures-sur-Yvette,
France} \\
$^{2)}$ {\sl Institute of Cosmology,  Department of Physics and Astronomy, Tufts University, Medford, MA 02155, USA}}
\maketitle

\begin{abstract}

The gravitational wave (GW) signals emitted by a network of cosmic
strings are reexamined in view of the possible formation of a network
of cosmic superstrings at the end of brane inflation.
The reconnection probability $p$ of intersecting fundamental or
Dirichlet strings might be much smaller than 1, and the properties of
the resulting string network may differ significantly from those of
ordinary strings (which have $p=1$). In addition, it
has been recently suggested that the typical length of newly formed
loops  may differ by a factor $\epsilon \ll 1$ from
its standard estimate. Here, we analyze the effects of the
two parameters $p$ and $\epsilon$ on the GW signatures of strings. We
consider both the GW bursts emitted from cusps of oscillating string
loops, which have been suggested as candidate sources for the
LIGO/VIRGO and LISA interferometers, and the stochastic GW background,
which may be detectable by pulsar timing observations. In both cases
we find that previously obtained results are \textit{quite robust}, at least when
the loop sizes are not suppressed by many orders of magnitude relative
to the standard scenario. We urge pulsar
observers to reanalyze a recently obtained 17-year combined data set
to see whether the large scatter exhibited by a fraction of the data
might be due to a transient GW burst activity of some sort, e.g.  to
a  near cusp event.
\end{abstract}

\section{Introduction}

Cosmic strings can be formed as linear defects at a symmetry breaking
phase transition in the early universe and can give rise to a variety
of observable phenomena at the present cosmic age. String formation,
evolution, and observational effects have been extensively studied in
the 1980's and 90's (for a review see
\cite{book,kibblehindmarsh}). This exploration was to a large degree
motivated by the string scenario of structure formation
\cite{Zeldovich,AV81a}, which requires strings of the grand
unification energy scale, $\eta \sim 10^{16}$ GeV, or $G\mu \sim
(\eta/M_p)^2 \sim 10^{-6}$. ( Here, $G$ is Newton's constant, $\mu$
is the string tension and $M_p\sim 10^{19}$ GeV is the Planck mass.
$G\mu$ is a dimensionless parameter
characterizing the gravitational interactions of strings.) This
scenario is now disfavored by the CMB observations, but strings of a
somewhat lower energy scale would be consistent with the data, and
their detection would of course be of great interest. The current CMB
bound on strings is \cite{Pogosian,Smoot} $G\mu < 6.1 \times
10^{-7}$. There is also a case of potential string detection with a
similar value of $G\mu$. Two nearly identical galaxies are observed at
angular separation of 1.9 arc sec, suggesting gravitational lensing by
a cosmic string with $G\mu\sim 4\times 10^{-7}$ \cite{Sazhin}.

Cosmic strings can also be detected through the gravitational wave
(GW) background produced by oscillating string loops
\cite{AV81b}. This background, which ranges over many decades in
frequency, has been extensively discussed in the literature
\cite{AV81b,Hogan,VV85,BB90,CA,Caldwell96}. The analysis of eight
years of millisecond pulsar timing observations has led to setting
rather stringent (95\% confidence level)
limits on the GW contribution to the cosmological
closure density $\Omega_{g} \equiv \rho_{g} / \rho_c \equiv 8 \pi G
\rho_{g}/ (3 H_0^2)$: $\Omega_{g} h^2 < 6 \times 10^{-8}$ according to
the original analysis \cite{KTR94}, or $\Omega_{g} h^2 < 9.3 \times
10^{-8}$ according to the Bayesian approach of \cite{MZVL96}. (Here,
$h \equiv H_0/(100 {\rm km/s/Mpc})$. Note that $h^2 \simeq (65/100)^2
\simeq 0.42$.)  As we shall review below, the corresponding bound on
$G\mu$ is $ c^{3/2}G\mu < 10^{-7} $, where $c$ denotes the (mean)
number of cusp events per oscillation period of a string loop.

Until recently, it appeared that the gravitational effects of strings
with $G\mu\ll 10^{-7}$ are too weak to be observable. However, it has
been shown in \cite{DV1,DV2} (following a suggestion in
\cite{BHV}) that GW bursts emitted from cusps of oscillating loops
should be detectable by LIGO and LISA interferometers for values of
$G\mu$ as low as $10^{-13}$.

During the last few years, there have been several important
developments that motivate us to reexamine the GW signatures of
strings. First, there has been a renewed interest in the possibility
\cite{Witten85} that fundamental strings of superstring theory may
have astronomical sizes and play the role of cosmic strings. In
particular, it has been argued \cite{Tye1,Tye2,Dvali} that fundamental (F)
and D-string networks can naturally be formed at the end of brane
inflation. In this scenario \cite{DvaliTye}, inflation is driven by
the attractive potential between a D-brane and an anti-D-brane, and
strings are produced when the branes eventually collide and
annihilate. The rather stringent requirements allowing for the stability
of such cosmic superstrings can be met in some scenarios
\cite{Dvali,polchinski,polchinski2}.
The predicted string tensions are
\cite{Tye1,Tye2,KKLMMT,polchinski} $10^{-11}\lesssim G\mu\lesssim
10^{-6}$ and appear to be comfortably within the range of
detectability by LIGO and LISA. However, the analysis of GW bursts in
\cite{DV1,DV2} may not be directly applicable in this case, since the
properties and evolution of F and D-string networks may differ in
significant ways from those of ``ordinary'' cosmic strings.

If both F and D-strings are produced, they can form an interconnected
network, in which F and D-strings join and separate at 3-way junctions
\cite{polchinski,Dvali}. Each junction joins an F-string, a D-string,
and a (1,1)-string, which is a bound state of F and
D. $(p,q)$-strings, which are bound states of $p$ F-strings and $q$
D-strings with $p,q>1$ can also be formed. The evolution of
FD-networks is similar to that of monopole-string $Z_3$ networks, in
which each monopole is attached to three strings. Simulations of $Z_3$
network evolution suggest \cite{VV87,Spergel} that the typical
distance between the monopoles scales with the cosmic time, $L\sim
\gamma t$, where the coefficient $\gamma$ depends on the rate of
energy loss by the network. But if the main energy loss mechanism is
gravitational radiation, as the case may be for an FD network, then it
has been argued in \cite{VV87} that the energy dissipation by the
network is rather inefficient, so it quickly comes to dominate the
universe.  If this picture is correct, then models predicting FD
networks are ruled out. It is conceivable, however, that the main
energy loss mechanism of FD networks is not GW emission, but chopping
off of small nets, similar to closed loop production by ordinary
strings. The negative verdict on this type of models can then be
avoided. This issue can only be resolved with the aid of new,
high-resolution numerical simulations of FD networks.

In this paper we shall focus on models where only one type of string
is formed. Still, the string evolution may differ from that of
ordinary strings, because the reconnection probability $p$ for
intersecting strings may be significantly smaller than 1. When
ordinary strings intersect, they always reconnect
\cite{Shellard,Matzner}. For intersecting F-strings, the reconnection
probability is suppressed by the string coupling, $g_s^2
<1$. Moreover, strings moving in a higher-dimensional bulk can avoid
intersection much more easily than strings in 3 dimensions
\cite{Dvali,Tye2}. The string propagation in the bulk is expected to
be restricted by bulk potentials, but the effective reconnection
probability can still be reduced by an order of magnitude or
so. Analysis in \cite{polchinski2} suggests reconnection probabilities
in the range
\beq
10^{-3}\lesssim p\lesssim 1
\label{p}
\eeq
for F-strings and
\beq
0.1\lesssim p\lesssim 1
\eeq
for D-strings. We thus need to analyze the effect of a reduced
reconnection probability on the GW burst statistics and on the
stochastic GW background.

Another interesting recent development has been the analysis by
Siemens and Olum \cite{Olum1} of the gravitational radiation from  counter-streaming wiggles
on long strings. They showed that this radiation is much less
efficient in damping the small-scale wiggles than originally
thought. This may result in much smaller sizes of closed loops produced
by the network \cite{Olum2} than previously assumed.
This effect can be quantified by the dimensionless  parameter
$\epsilon$ defined in Eq.~(\ref{epsilon}) below. Refs.~\cite{DV1,DV2} had
assumed (besides $p=1$), the ``standard'' value $\epsilon =1$, and we need to
see how a different value of $\epsilon$ might affect the GW burst statistics.

Last, but certainly not least on our list of recent developments is
the potential improvement in the sensitivity to a GW background of pulsar timing
observations. Indeed, these have been recently extended to a
17-year data set\cite{Lommen}. We shall discuss below what kind of
limits on $G\mu$ follow from this extended data set.

The main goal of the present paper is to analyze how the amplitude and
frequency of GW bursts from strings and the intensity of the
stochastic GW background depend on the parameter $\epsilon$ measuring the
 characteristic size of closed loops,
and on the string reconnection probability $p$. The paper is organized
as follows. In the next Section we outline the relevant features of string
evolution. GW bursts
from cusps are discussed in Section III. The stochastic background is discussed in Section IV,
where we also discuss bounds on $G\mu$ based on millisecond pulsar observations.
Our conclusions are summarized in Section V.

\section{String evolution}

\subsection{Standard scenario}

An evolving string network consists of two components: long strings
and sub-horizon closed loops. The long string component is
characterized by the following parameters: the coherence length
$\xi(t)$, defined as the distance beyond which the directions along
the string are uncorrelated, the average distance between the strings
$L(t)$, and the characteristic wavelength of the smallest wiggles on
long strings, $l_{wiggles}(t)$. The \textit{standard} picture of
cosmic string evolution (based on the assumptions
$p=1$ and $\epsilon =1$), asserts that (in units where the velocity of 
light is set to one)
\beq
L^{\rm st}(t) \sim \xi^{\rm st}(t) \sim t,
\label{L}
\eeq 
and
\beq
l_{wiggles}(t) \sim \alpha t,
\label{lwigg}
\eeq
where $\alpha$ is a constant whose value is specified below.

Strings move at relativistic speeds, and each long string intersects itself or
another long string about once every Hubble time $t$. As a result, one
or few large loops of
size $L^{\rm st}(t)\sim t$ are produced per Hubble time, which then shatter,
through multiple self-intersections, 
into a large number of small loops, whose size is comparable to the
wavelength of the wiggles (\ref{lwigg}). 
In other words, when loops are just formed they have a typical size
\beq
l(t) \sim \alpha t.
\label{l}
\eeq
This equation, referring to the typical size of \textit{just formed
loops}, will hold throughout this paper (including
in the ``non standard'' cases discussed below), and defines the 
meaning of the dimensionless loop-length parameter $\alpha$. 

Note that in order for the characteristic length of the network to
scale with cosmic time as in (\ref{L}),  
it is necessary for long strings to discharge a sizeable fraction of their length ($\sim t$ per Hubble volume
per Hubble time) in the form of closed loops. In other words, the number of loops formed
per Hubble volume per Hubble time, say $N_l$, is on the order of
\beq
N_l^{\rm st} \sim \frac{t}{l(t)} \sim \frac{1}{\alpha}.
\label{Nlst}
\eeq

The loops oscillate and lose their energy by gravitational radiation at the rate
\beq
d{\cal E}/dt \sim \Gamma G\mu^2,
\label{dE/dt}
\eeq
where $\Gamma\sim 50$ \cite{book} is a numerical coefficient. The lifetime of a loop
of length $l(t) \sim \alpha t$ and energy $ {\cal E} \sim \mu l \sim \mu \alpha t$ is
\beq
\tau\sim (\alpha/\Gamma G\mu)t.
\label{tau}
\eeq

The standard scenario assumes that the value of $\alpha$ is determined by the gravitational 
damping of small-scale wiggles. With the naive, old estimate of the damping, one finds
\cite{BB90}
\beq 
\alpha^{\rm st} \sim \Gamma G\mu.
\label{alpha}
\eeq
Then the lifetime of loops formed at time $t$ is $\tau\sim t$, so that, at any moment,
the number of loops per Hubble
volume is the same as the number of loops formed
per Hubble volume per Hubble time, as given by Eq.~(\ref{Nlst}).
Therefore the number density of loops in the \textit{standard} scenario is
\beq
n^{\rm st}(t)\sim \alpha_{\rm st}^{-1}t^{-3} \sim (\Gamma G \mu)^{-1}t^{-3} .
\label{n}
\eeq
In the following subsections we shall consider the modifications to the result
Eq.~(\ref{n}) of the standard scenario brought by two possible effects:
(i) a small reconnection probability $p \ll 1$, and (ii) a value of the loop-length parameter
$\alpha$ differing from the standard value Eq.~(\ref{alpha}) by being either smaller, or larger than it. We can quantify the effect (ii) with a new parameter
\beq
 \epsilon \equiv \alpha/ \alpha_{\rm st}  \equiv \alpha/ \Gamma G \mu.
\label{epsilon}
\eeq

Note again that we shall always define $\alpha$ by Eq.~(\ref{l}) (concerning the typical length of
newly formed loops), which therefore holds true both
in the standard scenario, and in the extended scenarios studied here. On the other hand,
Eq.~(\ref{alpha}) will cease to hold in the extended scenarios.

\subsection{Small reconnection probability: $p \ll 1$}

On sub-horizon scales, the strings are straightened out by the
expansion of the universe, and thus we expect, as before, that the
string coherence length is
\beq
\xi(t)\sim t.
\label{xip}
\eeq 
If the reconnection probability is $p\ll 1$, then one intersection per
Hubble time is not sufficient for ensuring the 
scaling of long strings. In order to have one reconnection, a long
string needs to have $\sim p^{-1}$ 
intersections per Hubble time. This means that the number of such
strings per Hubble volume should be $\sim p^{-1}$.

The typical inter-string distance $L(t)$ can be estimated from comparing two
different estimates of
the mean energy density in long strings, say $\rho_{long} $.
On the one hand, considering that a Hubble volume $\sim t^3$ contains
$\sim p^{-1}$ long strings yields $\rho_{long} \sim p^{-1} \mu
t/t^3 = \mu/(p t^2)$. On the other hand, the definition of $L$ is that
there should be $\sim 1$ string segment of length $L$
in a spatial volume $\sim L^3$, so that $\rho_{long} \sim \mu L/L^3 = 
\mu/L^2$. Equating the two estimates yields 
\beq
L(t) \sim p^{1/2} t,
\label{Lp}
\eeq
instead of  Eq.~(\ref{L}).

There have been some conflicting claims in the recent literature
regarding the values of $L(t)$ and $\xi(t)$ in the regime of $p\ll1$.
Jones, Stoica and Tye \cite{Tye2} find that $L(t) \sim \xi(t) \sim pt$.
We note, however, that they derive this result from the one-scale model 
\cite{Shellard04}, which assumes from the outset that $L(t) \sim \xi(t)$.
This assumption is not necessarily justified when $p\ll
1$. Sakellariadou \cite{Maria04} argues that $L(t)\sim \xi(t) \sim
p^{1/2}t$. Her analysis is based on numerical simulations of string
evolution in flat spacetime. Eq.~(\ref{Lp}) was indeed originally
obtained in such simulations \cite{Mairi}, but we are not aware of any
results indicating that $\xi(t)\sim L(t)$ in this regime.
On the contrary, preliminary results of a new, high-resolution flat-space
simulation indicate that $\xi(t)$ is substantially larger than $L(t)$ 
\cite{VOV}.

A definitive picture of string evolution with low intercommuting
probability can only be obtained from numerical simulations in an
expanding universe. Flat-space simulations do not include the
important effect of string stretching on sub-horizon scales. Simple
analytic estimates do not account for the possible effects of the
wiggles. For example, colliding wiggly strings may intersect at more
than one point, thus increasing the effective reconnection
probability. If the wiggliness of strings is increased at low $p$,
their velocity will be reduced, leading to a further decrease of
$L(t)$. String simulations with $p\ll 1$ will hopefully be performed
in the near future. For the time being, we shall assume that
Eqs.(\ref{xip}), (\ref{Lp}) give a reasonable approximation and use
them in the rest of the paper.

As before, a significant fraction of the total
string length within a Hubble volume $\sim  p^{-1} t $ should go into
loops each Hubble time, so that the number 
of loops formed per Hubble volume per Hubble time is now
\beq
N_l \sim \frac{p^{-1} t}{l(t)} \sim \frac{1}{p\alpha},
\label{Nl}
\eeq
i.e $p^{-1}$ larger than  the standard result (\ref{Nlst}).
In order to estimate the
 corresponding loop density at any moment, one must take into
 account the lifetime of these loops (which depends on the value of $\epsilon$).
  This will be discussed in the following subsections.

\subsection{Small loops: $\alpha\ll \Gamma G\mu$, i.e. $\epsilon \ll 1$}

As we already mentioned, recent analysis in \cite{Olum1} has shown
that the gravitational radiation from  counter-streaming wiggles
on long strings is much less efficient in damping the wiggles than
originally thought.  If indeed $\alpha$ is determined by gravitational
back-reaction, then the new analysis shows \cite{Olum2} that its value
is sensitive to the spectrum of small-scale wiggles, and is generally
much smaller than $\Gamma G\mu$. A simple model for the spectrum of
wiggles introduced in \cite{Olum2} yields 
\beq 
\alpha\sim (\Gamma G\mu)^n
\label{Olum}
\eeq
with $n=3/2$ in the radiation era and $n=5/2$ in the matter era.
In other words, the quantity (\ref{epsilon}) would be of order
$\epsilon \sim (\Gamma G\mu)^m$, with $m=n-1$, i.e.
$m=1/2$ in the radiation era and $m=3/2$ in the matter era.
As $\Gamma G\mu$ is observationally restricted to be
$\Gamma G\mu \lesssim 10^{-5}$, and might turn out to be $\ll 10^{-5}$,
this gives very small values of
$\alpha \lesssim 3 \times 10^{-8}$ and $\alpha\lesssim 3 \times 10^{-13}$,
and  $\epsilon \lesssim 3 \times 10^{-3}$ and $\epsilon \lesssim 3 \times 10^{-8}$,
for radiation and matter eras, respectively. In view of such possible drastic changes
in  orders of magnitude, it is important to assess the effect of $\epsilon \ll 1$ on
string network statistics and the corresponding GW observables.

As said above, for  $\alpha\ll \Gamma G\mu$, the number of loops produced per Hubble time per
Hubble volume is given by Eq.~(\ref{Nl}). However, to compute the number
of loops at any moment we must now take into account the fact that
 the lifetime of the loops (\ref{tau}) is $\tau\ll t$. Therefore, only a small
fraction of these loops will be present at any given time, $N'\sim (\tau/t)N \sim
1/(p\Gamma G\mu)$. The corresponding loop density is therefore
\beq
n (t)\sim {1\over{p\Gamma G\mu t^3}}.
\label{n<}
\eeq
Note that, compared to the standard result Eq.~(\ref{n}), the value of  $n(t)$ in scenarios extended by
the two parameters $p \ll 1$ and $\epsilon \ll 1$ is \textit{independent} of $\epsilon$, and exhibits
a simple dependence $\propto p^{-1}$ upon $p$.

\subsection{Large loops: $\alpha\gg \Gamma G\mu$, i.e. $\epsilon \gg 1$}

As mentionned above, the possibility $\alpha\ll \Gamma G\mu$, i.e. $\epsilon \ll 1$
will be our main focus in this paper, because of the recent findings of \cite{Olum1,Olum2}.
However, the opposite case $\alpha\gg \Gamma G\mu$, i.e. $\epsilon \gg 1$
cannot, at this stage, be dismissed. Indeed,
the spectrum of the wiggles is expected to be a power law decaying
towards shorter wavelengths \cite{Olum2}. This raises the possibility
that for sufficiently small wavelengths the wiggles may be too small
to have an effect on loop formation. The size of the loops will then
be determined by the dynamics of the network, and gravitational
back-reaction will play no role. 

The possibility of $\alpha\gg \Gamma G\mu$ has been discussed in \cite{book}.
We recall that the parameter $\alpha$ is defined by requiring that the typical
length of  \textit{newly  formed } loops is $\sim \alpha t_f$, where $t_f$ denotes
the time of formation. When $\alpha\gg \Gamma G\mu$, such loops will survive over many Hubble times,
and the loops extant at any given cosmological time $t$ will be obtained by integrating
over the loops formed on all formation times $t_f < t$. One finds that there is a
distribution of loops with sizes in the range $0 <l<\alpha t$, with the dominant contribution
to the number density, and to the GW burst rate, coming, at cosmological time $t$,
 from loops of typical size
\beq
l\sim \Gamma G\mu t.
\label{l2}
\eeq
These dominant loops at time $t$ were formed at the parametrically smaller typical time
 $t_f \sim (\Gamma G\mu/\alpha)t \ll t$. Their density at that (formation) time was
\beq
n_f \sim {1\over{p\alpha t_f^3}},
\eeq
and the density at time $t$ is 
\beq
n(t) \sim \left[{a(t_f)\over{a(t)}}\right]^3 n_f,
\label{n>}
\eeq
where $a(t)$ denotes the cosmological scale factor.
If we consider loops formed in the matter era, $t_f>t_{eq}$ (which will indeed be the most important for
GW observations), the factor $ (a(t_f)/a(t))^3= (t_f/t)^2 \sim   (\Gamma G\mu/\alpha)^2$, so that
the loop number density is
\beq
n(t)\sim {1\over{p\Gamma G\mu t^3}}.
\label{n>>}
\eeq
It is interesting to note that the final result (\ref{n>>}) for the loop density (in the matter era)
when $\epsilon \gg 1$ coincides
with the result (\ref{n<}) obtained in the opposite case $\epsilon \ll 1$, and that both
results are independent of $\epsilon$. Note, however, that the typical size of the loops
at time $t$ are different. In the case where $\epsilon \ll 1$ this typical size is
$l(t) \sim \alpha t = \epsilon \Gamma G\mu t$, while in the case $\epsilon \gg 1$
it is $l(t) \sim  \Gamma G\mu t$. In other words,  the case  $\epsilon > 1$ can be effectively
treated by taking the limit $\epsilon \to 1$ of  the case $\epsilon <
1$ (while keeping the effect of $p$). Another way to say this is to introduce the notion of \textit{effective} loop-length
parameter $\alpha_{eff}$, defined by writing that the typical size of the loops which dominate
the loop density at cosmic time $t$ is
\beq
l(t) \sim \alpha_{eff} t.
\label{alphaeff}
\eeq
Note that  Eq.~(\ref{alphaeff}) refers to the typical size of loops surviving at some
cosmic time $t$, while Eq.~(\ref{l}) referred  to the typical size of newly formed loops.
Correspondingly, we can define $\epsilon_{eff} \equiv \alpha_{eff}/ \Gamma G \mu$.
With this definition, one has $\alpha_{eff} = \alpha$ when $\alpha <\Gamma G \mu$
(i.e. $\epsilon_{eff} = \epsilon$ when $\epsilon < 1$), and
 $\alpha_{eff} = \Gamma G \mu$ when  $\alpha > \Gamma G \mu$
(i.e. $\epsilon_{eff} = 1$ when $\epsilon >1$). Note that $\epsilon_{eff}$ is
never greater than 1. [One could approximately write the link
 $\epsilon_{eff} = \epsilon/(1 + \epsilon)$.]

\subsection{This paper}

The bottom line is that the value of $\alpha$ is presently unknown. 
Numerical simulations of
string evolution \cite{BB90,AS} give loops that are too small to be
resolved, so only an upper bound on $\alpha$ can be obtained,
\beq
\alpha \lesssim 10^{-3}.
\eeq
The next generation of string simulations, which are now being
developed, are expected to improve this bound considerably. In this paper,
we shall use $\alpha$, or equivalently $\epsilon$ defined by Eq.~(\ref{epsilon}),
 as a free parameter and will allow it
to take values smaller, as well as greater than its standard value  $\alpha_{\rm st} = \Gamma G\mu$,
corresponding to $\epsilon_{\rm st} = 1$. Similarly, we shall
treat the reconnection probability $p$ as a free parameter and consider cases of both
high ($p=1$) and low ($p\ll 1$) reconnection probability.

It should be emphasized though that the parameters $p$ and $\alpha$ do not 
appear in the theory on equal footing. The reconnection probability $p$ and the string tension $\mu$ 
are true parameters, in the sense that they
can take different values in different cosmic string models (e.g., fundamental strings, 
D-strings, or ``ordinary'' strings). On the other hand, the parameter $\alpha$ only reflects
incompleteness of our understanding of string evolution and will eventually be determined, possibly as 
a function of $G\mu$ and $p$.\footnote{The parameter $c$, the average number of cusps per loop, which is
introduced in the next Section, is also not a true parameter and will hopefully be determined from
numerical simulations.}

\section{Gravitational wave bursts}

The main result of the GW burst analysis in Refs. \cite{DV1,DV2} is the expression for the
typical dimensionless amplitude of cusp-generated bursts, observed in an octave of
frequency around frequency $f$, that one can expect to
detect at a given occurrence rate ${\dot N}$ (say, one per year),
\beq
h_{\dot N}(f)\sim G\mu\alpha^{2/3}(ft_0)^{-1/3} g[y^{({\rm old})}]
\Theta(1 - \theta_m[\alpha,f,z_m(y^{({\rm old})})]).
\label{hold}
\eeq
Here, $t_0 \simeq 2/(3 H_0) \simeq 1.0 \times 10^{10} {\rm yr} \simeq 10^{7.5} {\rm s}$ is the present cosmic time,
\beq
y^{({\rm old})}({\dot N},f)=10^{-2}({\dot N}/c)t_0 \alpha^{8/3} (ft_0)^{2/3},
\label{oldy}
\eeq
$c$ is the average number of cusps per  period of loop oscillation, and the superscript (old)
refers to the fact that Eq.(\ref{oldy}) was derived using the old (standard) string evolution
scenario with $\alpha\sim\Gamma G\mu$. The function $g[y]$ in (\ref{hold}) is given by
\beq
g[y]=y^{-1/3}(1+y)^{-13/33}(1+y/y_{eq})^{3/11}.
\label{g}
\eeq
This is an interpolating function which represents the power-law behavior of $h_{\dot N}$ in
three different regimes: $y\lesssim 1$, $1\lesssim y\lesssim y_{eq}$, and $y\gtrsim y_{eq}$,
where $y_{eq}=z_{eq}^{11/6}$ and $z_{eq}\simeq 10^{3.94}$ is the redshift of equal matter and radiation densities.
The three regimes correspond to loops radiating, respectively, at $z\lesssim 1$, 
$1\lesssim z\lesssim z_{eq}$, and $z>z_{eq}$.
Finally, the last factor $\Theta$ in Eq.(\ref{h}) involves Heaviside's step function $\Theta$, with the
argument $1 - \theta_m$ obtained by inserting  Eq.(\ref{oldy}) into the function
\beq
z_m(y) =y^{1/3}(1+y)^{7/33}(1+y/y_{eq})^{-3/11},
\eeq
and then by inserting the result into the function
\beq
\theta_m(\alpha,f,z) = (\alpha f t_0)^{-1/3} (1+z)^{1/6} (1+z/z_{eq})^{1/6}.
\label{theta}
\eeq
Physically, the quantity $\theta_m(f)$ is related to the (integer) mode number $m$ of the Fourier
decomposition of the gravitational radiation emitted by a loop by $ \vert m \vert \sim (\theta_m(f))^{-3} $.
Note also the link $\vert m \vert  \sim (1+z) f l$ \cite{DV2}, where $l$ denotes the length of the loop.
The $\Theta$-function factor serves the purpose of restricting the burst signals to the values
$\theta_m \leq 1$, corresponding to $ \vert m \vert  \geq 1$.
In view of Eq.~(\ref{theta}), this step function is  equal to one when the product
 $\alpha f t_0$ is larger than 1 and $z \lesssim 1$.
  This case covers many cases of physical interest, especially when considering
 GW frequencies in the LIGO or LISA bands. This is why, in most of our analytical discussion below
 we shall only consider the other factors in Eq.~(\ref{hold}). However, we shall include the effect of this cut-off factor in our plots below,
 and we shall see that is plays a crucial role for the GW frequencies of relevance in pulsar timing
 observations, and that it also starts playing an important role for LIGO and LISA
 signals, when $\epsilon$ gets much smaller than 1.

The log-log plot of $h_{\dot N}$ as a function of $G\mu$, calculated for the standard scenario, and in absence
of the cut-off brought by the $\Theta$-function factor, is made up of
three straight lines representing the three different regimes mentioned above. Now we would like to find
out how these lines are modified when $\epsilon\not= 1$ and/or $p< 1$.
By looking at the derivation of Eqs.(\ref{hold})-(\ref{g})
in Refs.\cite{DV1,DV2} one finds that the explicit occurrences of the notation $\alpha$ concerned two different aspects of
a string network: (i) either  $\alpha$ was used to parametrize the typical
size of a loop, in the sense of Eq.~(\ref{l}), or better of Eq.~(\ref{alphaeff}) as one is interested in the
typical size of a loop at cosmic time $t$,  or, (ii) $\alpha$ entered as a factor in the loop density, written as
$n(t)\sim \alpha^{-1}t^{-3}$. The first usage of   $\alpha$ is consistent with the definition  Eq.~(\ref{l}) of $\alpha$
in the extended scenarios considered here  in the most interesting case where $\epsilon <1$; as said above,
we shall treat the opposite case by replacing $\alpha \to {\alpha}_{eff}$, \textit{i.e.} by
taking the $\epsilon \to 1$ limit. But then, if $\alpha$ is so defined, the second
usage must be corrected for because the loop density in extended scenarios is not given by $n(t)\sim \alpha^{-1}t^{-3}$,
but rather by  Eqs.~(\ref{n<}) and (\ref{n>}) (the latter becoming Eq.~(\ref{n>>}), which is the same as (\ref{n<}) in the
matter era case). In other words, we just need to correct (at least when $\epsilon < 1$),
the loop density used in Refs.\cite{DV1,DV2}  by the factor $\alpha/ p \Gamma G \mu = \epsilon/p$.
An easy way to accomplish this is to note that the modified
loop density can be accounted for by adjusting the value of the number of cusp events per loop
oscillation, $c$. Specifically,
for $\epsilon\lesssim 1$ we need to make a replacement
\beq
c\to {c\epsilon\over{p}}.
\label{newc}
\eeq
Alternatively, as $c$ always enters the final results in the combination $ {\dot N}/c$, we could
account for the modified loop density by the replacement
\beq
{\dot N} \to { {\dot N} p \over{\epsilon}}.
\eeq
As a result, Eqs.(\ref{hold}) and (\ref{g}) remain unchanged, while Eq.(\ref{oldy}) is replaced by
\beq
y^{({\rm new})}({\dot N},f)=10^{-2}{p\epsilon^{5/3}\over{c}}({\dot N}t_0) \alpha_0^{8/3} (ft_0)^{2/3}
= p\epsilon^{5/3} y^{({\rm old})}({\dot N},f, \alpha_0) ,
\label{newy}
\eeq
where $\alpha_0\equiv \Gamma G\mu$, i.e. $\alpha_0$ denotes what was denoted $\alpha_{\rm st}$ above.
 To keep track of the dependence on $\epsilon$ (at a fixed $G\mu$),
we shall also rewrite Eq.(\ref{hold}) in terms of $\epsilon$ and $\alpha_0$,
\beq
h_{\dot N}(f)\sim {\Gamma}^{-1}\epsilon^{2/3} \alpha_0^{5/3}(ft_0)^{-1/3} g[y^{({\rm new})}]
\Theta(1 - \theta_m[\alpha,f,z_m(y^{({\rm new})})]) .
\label{h}
\eeq
The function $g[y]$ has the successive power-law behaviours $g[y] \propto y^n$ with
$n= -1/3$ for $y\lesssim 1$, $n= -8/11$ for $1\lesssim y\lesssim y_{eq}$, and
$n=-5/11$ for $y\gtrsim y_{eq}$. In view of the scaling $y^{({\rm new})} \propto p\epsilon^{5/3}$,
Eq.~(\ref{newy}), we see from Eq.~(\ref{h}) that $h_{\dot N}(f)$ will have successive power-law
scalings with $p$ and $\epsilon$ of the form
$h_{\dot N}(f) \propto \epsilon^{2/3} (p\epsilon^{5/3})^n = p^n \epsilon^{(2+5n)/3}$.
Therefore the value of $h_{\dot N}(f)$ for $p<1$, $\epsilon<1$ can be obtained from the
corresponding value, for the same values of $f, G \mu, {\dot N}$ and $c$, and
 for $p=\epsilon=1$, by multiplying it with the following factors: $p^{-1/3}\epsilon^{1/9}$
for $y\lesssim 1$, $p^{-8/11}\epsilon^{-6/11}$ for $1\lesssim y\lesssim y_{eq}$, and $p^{-5/11}\epsilon^{-1/11}$
for $y\gtrsim y_{eq}$.

The qualitative effect of varying $p$ and $\epsilon$ is now easy to understand. The graph of $h_{\dot N}$ {\it vs.}
$G\mu$ has a zigzag shape, with a rising line on the left, a short declining segment in the middle, and
another rising line on the right. When $p$ is lowered, at fixed $G\mu$ and $\epsilon$, all three lines move up,
with the central segment moving somewhat more than the right line, and the right line somewhat more than the left line.
When $\epsilon$ is lowered at constant $G\mu$ and $p$, the left line moves down, while the central segment and the 
right line move up. The displacements of the three lines in this case are very unequal. The displacement of the left 
and right lines is small, unless $\epsilon$ changes by 5 orders of magnitude or more, while the displacement of the
middle segment is quite noticeable, even if $\epsilon$ is changed by only one or two orders of magnitude.
It can be easily seen that the local maximum and minimum of the curve $h_{\dot N}(G\mu)$ are both shifted to the 
right if $\epsilon$ is decreased (see below). When $\epsilon$ and $p$ are changed independently of $G\mu$, the slopes of all
three lines remain unchanged. This would not be the case if, for example, $\epsilon$ were a function of $G\mu$, 
as in Eq.(\ref{Olum}).

A complementary way to qualitatively grasp the effect of $p<1$ and $\epsilon<1$ on the plot of  $h_{\dot N}(f)$
 (considered for a fixed value of $f$) versus $G \mu$ is the following. In Eq.~(\ref{h}) above the dependence of
 $h_{\dot N}(f)$ on  $G \mu$ comes both through the prefactor $\alpha_0^{5/3}$ and through the
 dependence of $y^{({\rm new})}$ upon $\alpha_0$ appearing in Eq.~(\ref{newy}), namely
  $y^{({\rm new})}  \propto \alpha_0^{8/3}$.  We can, however, have a fast grasp at the location and height of the
  \textit{extrema} of the zig-zagged graph of $h_{\dot N}$ {\it vs.} $G\mu$ by using the inverse dependence
  $\alpha_0 \propto  ( y^{({\rm new})} )^{3/8}$ to reexpress $h_{\dot N}$ as a function of $y^{({\rm new})}$,
 instead of $\alpha_0$, or $G \mu = \alpha_0/50$. Keeping all the factors, this leads to
\beq
h_{\dot N}(f)\sim \epsilon^{-3/8}  p^{-5/8} H_1(f ,t_0,c, \dot N) G(y^{({\rm new})} ),
\label{h'}
\eeq
where we have introduced the auxiliary functions
\beq
 H_1(f ,t_0,c, \dot N) = \frac{1}{\Gamma}  {\left[ \frac{10^2 c}{\dot N t_0} \right]}^{5/8} (f t_0)^{-3/4}
 \eeq
 and
\beq
G[y] \equiv y^{5/8} g[y] =y^{7/24}(1+y)^{-13/33}(1+y/y_{eq})^{3/11}.
\label{G}
\eeq
The function $G(y)$ has a maximum at $y\sim1$, and a minimum at $y \sim y_{eq}$. These being
numerically fixed values, the
result Eq.~(\ref{h'}) above immediately shows that the \textit{heights} of the extrema of the plot
$h_{\dot N}$ {\it vs.} $G\mu$ depend on $\epsilon$ and $p$ only through the simple prefactor
in (\ref{h'}). Namely, the heights of the extrema are proportional to
\beq
h^{\rm extrema} \propto \epsilon^{-3/8}  p^{-5/8}.
\eeq
Interestingly, the effect of $\epsilon <1$ or $p<1$ is always to \textit{increase} the extremal values
of the plot $h_{\dot N}$ {\it vs.} $G\mu$. As for the \textit{locations} of these extrema on the $G \mu$ axis,
they are given by inverting the relation (\ref{newy}) linking $y$ to $ G\mu$. This yields (modulo
factors depending only on $c, \dot N, f$ and $t_0$)
\beq
G \mu_* \propto \epsilon^{-5/8}  p^{-3/8} y_*^{3/8}
\label{Gmu}
\eeq
where $y_*$ denotes the location of an  extremum on the $y$ axis ,i.e. $y_* \sim 1$ or
 $y_* \sim y_{eq}$. These being numerically fixed values, we see that the \textit{locations} on
 the $G \mu$ axis of the extrema of $h_{\dot N}(f)$ vary with $\epsilon$ and $p$ simply through the
 prefactor $\epsilon^{-5/8}  p^{-3/8}$ in Eq.~(\ref{Gmu}). This factor is always larger than 1 when
$\epsilon <1$ or $p<1$. In summary, the effects of $\epsilon <1$ and $p <1$ on the zig-zagged
graph of $h_{\dot N}$ {\it vs.} $G\mu$ is to move it \textit{up} by a factor $\epsilon^{-3/8}  p^{-5/8}$,
and \textit{right} by a factor $ \epsilon^{-5/8}  p^{-3/8}$. It is easily checked that this simple
behaviour is compatible with the more detailed results above about the ``motion'' with $\epsilon$
and $p$ of the three lines making up the zig-zagged plot $h_{\dot N} (G\mu)$, when taking into account
the successive logarithmic slopes ($+7/9, -3/11, + 5/11$ \cite{DV2}) of  $h_{\dot N}(f)$ versus $G\mu$.

How do these modifications affect the detectability of the GW bursts from cusps? A smaller value of $p$ can only 
increase $h_{\dot N}$, and therefore improves the detectability. The improvement is moderate when
considering a given value of $ G \mu$: for $p\sim 10^{-3}$,
we gain one order of magnitude in $h$. The improvements in the detectability of GW burst signals in the
LIGO and LISA detectors brought by reducing the value of $p$ (from 1 to $10^{-3}$, in steps of $10^{-1}$)
is illustrated in Figs. 1 and 2.

\begin{figure}
\begin{center}
\hspace{-0.8cm}
\epsfig{file=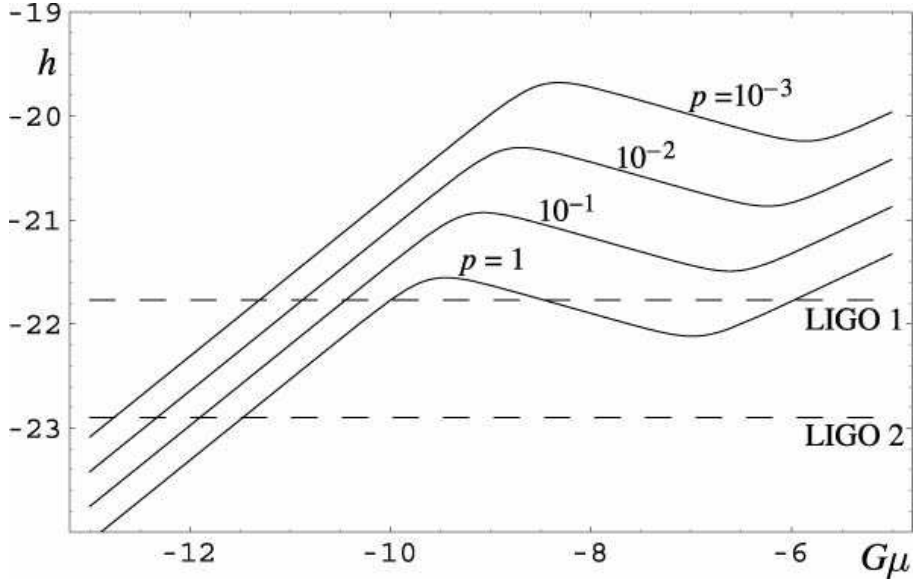,width=12cm}
\medskip
\caption{\sl Effect of a reconnection probability $10^{-3} \leq p \leq 1$ on the
gravitational wave amplitude of bursts emitted by cosmic string cusps
in the LIGO/VIRGO frequency band ($f_{ligo} =150$ Hz), as a function of the string tension
parameter $ G \, \mu$ (in a base-$ 10$ log-log plot). Here, as in the following figures, the
average number of cusps per loop oscillation is assumed to be $c=1$.
The horizontal dashed lines indicate the one sigma noise levels (after optimal
filtering) of LIGO~1 (initial detector) and LIGO~2 (advanced configuration). }
\label{Fig1}
\end{center}
\end{figure}

The horizontal dashed lines in these figures indicate the one sigma noise levels,
after optimal filtering, for detecting GW bursts in the corresponding detectors. In the
``LIGO'' figures, the  upper horizontal line corresponds to the initial sensitivity, LIGO 1, or equivalently VIRGO,
while the lower line corresponds to the planned advanced configuration LIGO 2.
For details of the signal to noise analysis, see \cite{DV2}.

\begin{figure}
\begin{center}
\hspace{-0.8cm}
\epsfig{file=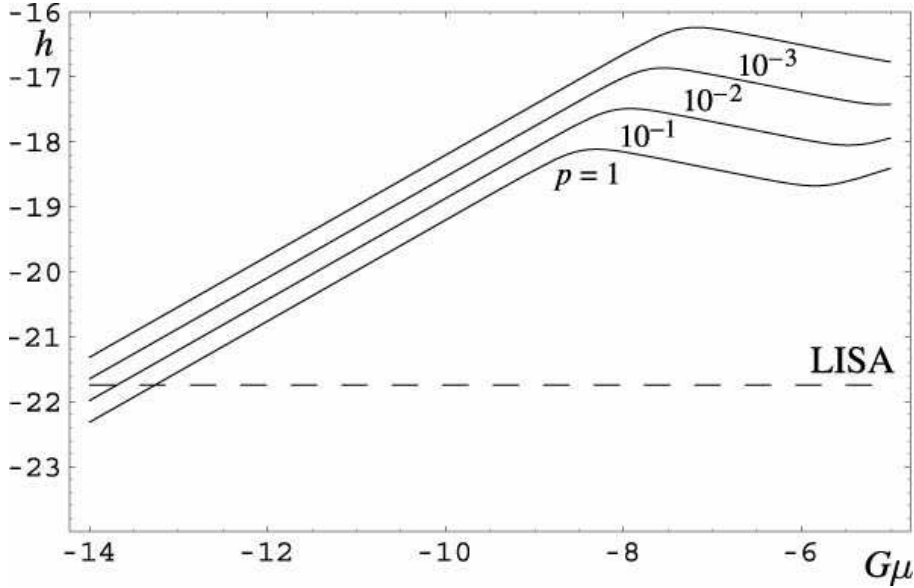,width=12cm}
\medskip
\caption{\sl Effect of a reconnection probability $10^{-3} \leq p \leq 1$ on the
gravitational wave amplitude of bursts emitted by cosmic string cusps
in the LISA frequency band ($f_{lisa} =3.88 \times 10^{-3}$ Hz) , as a function of the string tension
parameter $ G \, \mu$ (in a base-$ 10$ log-log plot).
The horizontal dashed line indicates the one sigma noise level (after optimal
filtering) of the LISA detector. }
\label{Fig2}
\end{center}
\end{figure}

Turning to the effect of $\epsilon <1$,  we find that the burst amplitudes $h_{\dot N}(f)$ are very weakly
affected by a decrease of $\epsilon$ by a few orders of magnitude. This follows from the
fact that the limiting sensitivities of LIGO and LISA correspond to the regime $y<1$, where the amplitude $h$ is 
only lowered by a factor $\epsilon^{1/9}$. From this factor it would seem that it is only in the case
where $\epsilon \lesssim 10^{-10}$ that detectability by LIGO or LISA will be significantly
affected. However, when $\epsilon$ gets that small the $\Theta$-function factor in Eq.~(\ref{h}) starts
playing an important role even at LIGO or LISA frequencies, especially when considering the types
of values of the string tension, $ G \mu \sim 10^{-10}$, suggested by recent stringy implementations
of brane inflation \cite{KKLMMT,polchinski}. Indeed, the crucial numerical factor in the $\Theta$-function
cut-off is the product
$\alpha f t_0 = \epsilon \Gamma G \mu f t_0 \sim 10^{9.2} \epsilon (G \mu/ 10^{-10}) (f/ {\rm Hz})$
entering $\theta_m$, Eq.~(\ref{theta}).
Therefore, when $ G \mu \sim 10^{-10}$, the cut-off brought by the $\Theta$ function starts
significantly affecting the LIGO signal ($f_{LIGO} \sim 100$ Hz) when $\epsilon \lesssim10^{-11}$,
and the LISA one ($f_{LISA} \sim 10^{-2}$ Hz) when $\epsilon \lesssim10^{-7}$.
The effect of $\epsilon <1$ on the detectability of GW bursts by LIGO and LISA is illustrated
in Figs. 3 and 4.

\begin{figure}
\begin{center}
\hspace{-0.8cm}
\epsfig{file=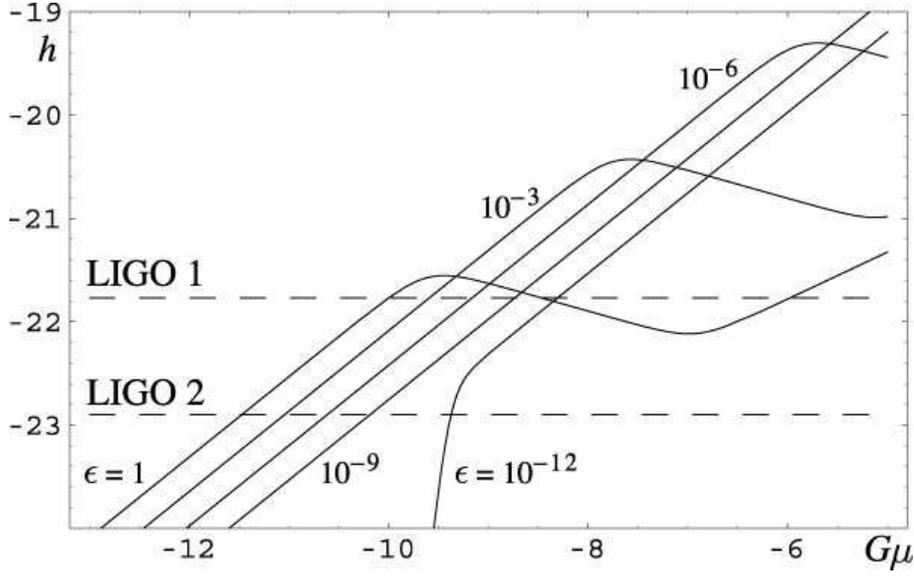,width=12cm}
\medskip
\caption{\sl Effect of a smaller fractional loop-length parameter
$10^{-12} \leq \epsilon \equiv \alpha/(50 G \mu) \leq 1$ on the
gravitational wave amplitude of bursts emitted by cosmic string cusps
in the LIGO/VIRGO frequency band ($f_{ligo} =150$ Hz), as a function of the string tension
parameter $ G \, \mu$ (in a base-$ 10$ log-log plot).
The horizontal dashed lines indicate the one sigma noise levels (after optimal
filtering) of LIGO~1 (initial detector) and LIGO~2 (advanced configuration). }
\label{Fig3}
\end{center}
\end{figure}

\begin{figure}
\begin{center}
\hspace{-0.8cm}
\epsfig{file=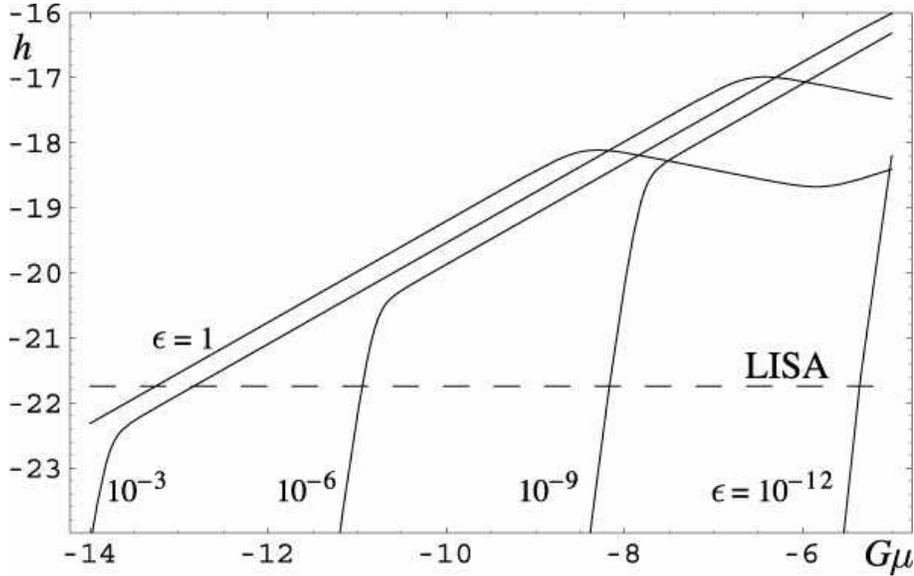,width=12cm}
\medskip
\caption{\sl Effect of a smaller fractional loop-length parameter
$10^{-12} \leq \epsilon \equiv \alpha/(50 G \mu) \leq 1$ on the
gravitational wave amplitude of bursts emitted by cosmic string cusps
in the LISA frequency band ($f_{lisa} =3.88 \times 10^{-3}$ Hz) , as a function of the string tension
parameter $ G \, \mu$ (in a base-$ 10$ log-log plot).
The horizontal dashed line indicates the one sigma noise level (after optimal
filtering) of the LISA detector. }
\label{Fig4}
\end{center}
\end{figure}

Let us now briefly discuss the case $\epsilon >1$. In that case, the effective value of $\alpha$ when
 $\alpha$ is used to parametrize the typical size of a loop at cosmic time $t$ is
 $\alpha_{eff} = \Gamma G \mu$, corresponding to $\epsilon_{eff} =1$, while
the loop density is
generally given by  Eq.(\ref{n>}). The latter result takes different explicit forms, according
to whether one considers the cases of loops formed in the matter era
(then one gets Eq.(\ref{n>>})), loops formed in the radiation era and decaying in matter era, or loops both
formed and decaying in the radiation era (see \cite{book}). For loops formed in the matter era, which correspond 
to the most relevant part of the graph in terms of  observability, the loop density Eq.(\ref{n>>}) coincides with
the $\epsilon =1$ limit of the case $\epsilon <1, p<1$ considered above. Therefore, the plot of $h$ \textit{vs.} $G \mu$
is obtained from the discussion above by keeping the effect of $p <1$, while setting $\epsilon =1$. In that case,
we see that the net effect is to increase the detectability of GW bursts.

The overall conclusion is that the results of our previous work
\cite{DV1,DV2} are quite robust\footnote{Let us also mention that Ref.~\cite{SO03} has
shown that the waveform of GW bursts from cusps derived in \cite{DV1,DV2} is robust against
the presence of small-scale wiggles on the strings.}
 against the inclusion of the
modifications parametrized by $p<1$ and by $\epsilon$ (both when $
\epsilon <1$ and $\epsilon >1$), at least when $\epsilon \gtrsim
10^{-11}$ in the case of LIGO, or $\epsilon \gtrsim 10^{-7}$ in the
case of LISA.  In particular, it is notable that a smaller
reconnection probability 
can only increase the
detectability of cosmic superstrings, and that even a very small
$\epsilon$ leads to a vast range of detectability for the GW bursts
emitted by a network of strings. However, smaller values of $\epsilon$
might lead to cutting off the burst signal in GW detectors, when the
string tension is $G \mu \sim 10^{-10}$ or less. See Figs. 1-4.

\section{Stochastic Gravitational Wave Background}

We now consider the stochastic GW background produced by oscillating string loops.
It was shown in \cite{DV1,DV2} that this background appears, in general, as the superposition of
occasional (non Gaussian) bursts, on top of
a nearly Gaussian ``confusion noise''  $h^2_{confusion}(f)$, made of overlapping bursts.
For some (rather large) values of the effective loop-length parameter $\alpha_{eff}$, Eq.~(\ref{alphaeff}), namely
$\alpha_{eff} \sim 10^{-5}$ in the standard case, the individual burst events detectable during
a typical pulsar-timing observation time scale $ T \sim 10$ yr have an amplitude which might
be comparable to the confusion background. This raises subtle issues about the detectability of such
a mixed Gaussian-non-Gaussian background. In the  discussion
 of this section, we shall, however, restrict ourselves to considering the simpler case of
relatively small values of the effective loop-length parameter where the individual, non-overlapping,
bursts are negligible compared to the background ``confusion noise''.  Note, also, that, even in this case,
Refs. \cite{DV1,DV2} found that the quantity  usually considered in the cosmic string
literature, namely the ``rms'' noise $h^2_{rms}(f)$, averaged
over all bursts, was not necessarily a good estimate of the observationally relevant confusion noise,
$h^2_{confusion}(f)$. Indeed,  $h^2_{rms}(f)$ includes, contrary to $h^2_{confusion}(f)$,
the time-average contribution of rare,
intense bursts, which are not relevant to a pulsar experiment of limited duration.

The confusion noise can be written as the following integral over the redshift $z$
(using Eq.~(6.17) of \cite{DV2} with the replacement Eq.~(\ref{newc}) above)

\beq
h^2_{confusion}(f) = \int \frac{dz}{z} n(f,z) h^2(f,z) \Theta[n(f,z) -1],
\label{hconfusion}
\eeq
where
\beq
n(f,z) = 10^2 \frac{c \,\epsilon}{p} (f t_0)^{-5/3} {\alpha}^{-8/3} \varphi_n(z) {\cal C}(z) ,
\label{n(fz)}
\eeq
is the number of cusp events  (in a time window $\sim f^{-1}$ and
 around redshift $z$) generating GW bursts around frequency $f$. Here
\beq
\varphi_n(z) =  z^3 (1+z)^{-7/6}(1+z/z_{eq})^{11/6},
\eeq
\beq
{\cal C}[z] = 1 + 9 z/(z + z_{eq}),
\eeq
and
\beq
h(f,z) = G \mu \alpha^{2/3} (f t_0)^{-1/3} \varphi_h(z) \Theta(1 - \theta_m[\alpha,f,z] ),
\label{hfz}
\eeq
with
\beq
\varphi_h(z) =  z^{-1} (1+z)^{-1/3}(1+z/z_{eq})^{-1/3}.
\eeq
The factor ${\cal C}[z] $ interpolates between 1 in the matter era and 10 in the radiation era.
It is incorporated to refine, in the standard case at least, the order of magnitude
estimate Eq.~(\ref{n<}) of the loop density.  The quantity $\theta_m[\alpha,f,z]$ entering the
step function in the burst amplitude $h(f,z)$ is that defined in Eq.(\ref{theta}) above.
As above the step function $\Theta(1 - \theta_m[\alpha,f,z] )$ cuts off the Fourier
components that would correspond to a mode number $ m \lesssim 1$. Note the presence
of a second step function, $\Theta[n(f,z) -1]$ in $h^2_{confusion}(f)$, which (approximately) limits
the integral over the redshift $z$ to the \textit{overlapping} bursts. Indeed, the ``confusion noise''
 Eq.(\ref{hconfusion}) is obtained by subtracting from the ``rms noise''
$h^2_{rms}(f) = {\int}_{0}^{\infty} (dz/z) n(f,z) h^2(f,z)$ the signals of low redshift
${\int}_{0}^{z_c(f)} (dz/z) n(f,z) h^2(f,z)$ with $n(f,z_c(f)) \sim 1$. As shown in \cite{DV2}, the
latter signals can be intense, but they are rare and non-overlapping, their total number (in a time
window $\sim f^{-1}$) being
$\sim {\int}_{0}^{z_c(f)} (dz/z) n(f,z) \sim n(f,z_c(f)) \sim 1$. (We use here the fact that $n(f,z)$ is
a monotonically increasing, power-law-type, function of $z$.)

Finally, we associate to the dimensionless squared GW amplitude (per octave of frequency)
$h^2_{confusion}(f)$ its  energy density per octave of frequency,
 $f d\rho_g/df \sim [ 2\pi f h_{confusion}(f)]^2/(16\pi G) \sim (\pi/4G) f^2 h^2_{confusion}(f)$, and
 the corresponding fractional contribution
(per octave of frequency) of the
confusion GW noise to the cosmological closure density,
$\Omega_g(f) = (f/\rho_c)(d\rho_g/df)$, where $\rho_c\sim 1/(6\pi Gt_0^2)$
(see  Eq.~(6.20) of \cite{DV2})
\beq
\Omega_g^{confusion}(f) \sim \frac{3 \pi^2}{2} (f t_0)^2 h^2_{confusion}(f).
\label{omega=h2}
\eeq

This quantity is plotted, for a  pulsar timing frequency $f_{psr} \sim 0.1 {\rm yr}^{-1}$
corresponding to a typical $\sim 10$ yr observational window, as a function of $G\mu$,
and for various values of $p$ and $\epsilon$, in Figs. 5 and 6. The horizontal lines in
these figures correspond to various realized, or planned, pulsar timing experiments.
The upper (continuous) line corresponds to the  (95\%  confidence level) upper limit
$\Omega_g h^2 < 6 \times 10^{-8}$ derived in \cite{KTR94} from 8 years of high-precision
timing of two millisecond pulsars: PSR  1855+09 and PSR 1937+21.
 [Note that a Bayesian reanalysis of the data of \cite{KTR94}
gave, under the choice of ``Jeffrey's prior'', the slightly less stringent limit
$\Omega_g h^2 < 9.3 \times 10^{-8}$ \cite{MZVL96}. Note also that, consistently
with our use of $H_0 \simeq 65$ km/s/Mpc and  $t_0 \simeq 1.0 \times 10^{10}$ yr,
we have $h^2 \simeq (65/100)^2 \simeq 0.42$]

 Recently, the data set for these
two pulsars has been extended to a 17-year continuous span by piecing together
data obtained from three different observing projects \cite{Lommen}. This is the
first realization of the concept of Pulsar Timing Array  (PTA) \cite{Backer1,Backer2}.
However, the upper limit
on $\Omega_g h^2$ that one can deduce from this 17-year combined data set is unclear to us.
On the one hand, \cite{Lommen} computes two widely different upper limits by using
two different approaches.  A Neyman-Pearson test leads, according to \cite{Lommen},
to a 95\% confidence level limit of only $\Omega_g h^2 < 2.8 \times 10^{-6}$, which is
much less stringent than the limit obtained in \cite{KTR94} from 8 years of data. In view of
this surprising result, Ref. \cite{Lommen} then resorted to a rather coarse estimate of an upper limit,
based only on saying that ``the largest amplitude sinusoid that one could conceivably
fit to the PSR 1855+09 data'' is 3 $\mu$s, for a frequency $f = 1/(17)$ yr.  This led to the upper limit
$\Omega_g h^2 < 2 \times 10^{-9}$, which is now more than ten times more stringent than the limit
based on 8 years of data. On the other hand, a  look at the PSR  1855+09 residuals reported in \cite{Lommen}
shows that, during an  intermediate period of $\sim 5$yr corresponding to the Green Bank data, there were
 residuals reaching the  $\sim 30 \mu$s level, \textit{i.e.} a much larger level than the $\sim 3 \mu$s level
 typical of the (pre- and post-upgrade) Arecibo data. This large ``activity''  of the pulsar data points during the
 Green-Bank-only period is responsible for the $\Omega_g h^2 < 2.8 \times 10^{-6}$ Neyman-Pearson-test limit.
 Apparently, the author of \cite{Lommen} seems to favor  the tighter limit $\Omega_g h^2 < 2 \times 10^{-9}$,
 obtained without using any statistical reasoning, more than the Neyman-Pearson-test one $\Omega_g h^2 < 2.8 \times 10^{-6}$.
 Personally, in view of the  ``large activity'' exhibited by the Green Bank data,
 we are not convinced that the level $\Omega_g h^2 = 2 \times 10^{-9}$ can be considered as
 a real upper limit on $\Omega_g h^2$. Pending a more complete statistical analysis of the present 17-year
 combined pulsar data, we can only consider this level as the potential sensitivity level of such
 an extended data set. Accordingly, we have represented the level $\Omega_g h^2 = 2 \times 10^{-9}$
 as a  \textit{dashed} line in Figs. 5, 6.

\begin{figure}
\begin{center}
\hspace{-0.8cm}
\epsfig{file=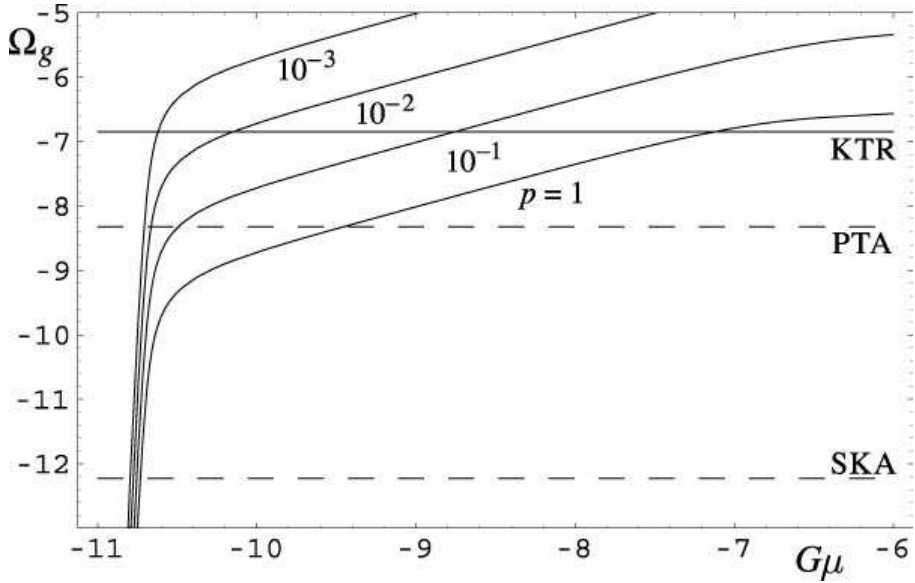,width=12cm}
\medskip
\caption{\sl Effect of a reconnection probability $10^{-3} \leq p \leq 1$ on the
fractional contribution $\Omega_g(f_{psr})$ (around the frequency $f_{psr} \sim 1/(10 {\rm yr})$) to the cosmological closure density of the
stochastic GW noise due to overlapping GW bursts emitted by a network of strings.
[Base-$ 10$ log-log plot.] The upper (solid) horizontal line indicates the upper limit
$\Omega_g  < 6  h^{-2}\times 10^{-8}$ derived from 8 years of high-precision
timing of two millisecond pulsars: PSR  1855+09 and PSR 1937+21.
The middle (dashed) horizontal line
indicates the potential sensitivity of  17 years of high-precision timing of PSR  1855+09
(see  text). The lower (dashed) horizontal line indicates the
expected sensitivity from the timing of the set of pulsars to be hopefully detected
by a square-kilometer-array of radio telescopes. }
\label{Fig5}
\end{center}
\end{figure}

 Actually, we wish to suggest, with due reserve,  that the large scatter recorded in the Green Bank data
 set might be due to the real effect of a transient GW burst activity of some sort, e.g.  to a
  near ($z \ll 1$) or rare ($c \ll 1$) cusp event (indeed, there is no evidence of a steady stochastic red noise in
 the Arecibo data). When looking not only at the
 PSR  1855+09 data, but also at the PSR 1937+21 ones, one notices that, after fitting out a cubic
 term $\propto {\ddot P}$ in the residuals, there remains some larger-than-usual activity, at the $ 4 \mu$s
 level. It is clearly too early to use such data to fit for possible string-loop parameters, but we urge
 the observers not to dismiss this interesting possibility.

 Finally, looking ahead to the realization of the project of the Square Kilometer Array
 radio telescope \cite{SKA}, and of its consequent pulsar timing array, one can ultimately
 hope to reach, through pulsar timing, the sensitivity level
 $\Omega_g h^2 =  10^{-12.6}$ at a frequency $f_{psr} \sim 0.1 {\rm yr}^{-1}$ \cite{SKA}.
 This ultimate sensitivity level is indicated as the lowest dashed line in Figs. 5, 6.

 Figs. 5, 6 illustrate the effects of $p$ and $\epsilon$ on the detectability of the
 stochastic GW background generated by a cosmological network of (super)strings.
 This background is essentially made of the superposition of overlapping burst
 signals. Therefore we expect that the effects of $p$ and $\epsilon$ discussed
 above on the case of individual bursts will somehow extend to this ``confusion noise''.
 Indeed, Fig. 5 shows that decreasing $p$ increase the signal, and therefore
 improves its detectability by pulsar timing experiments.  Concerning the effect
 of decreasing $\epsilon$ the situation is a bit different from what happened in
 the case of individual bursts in the LIGO or LISA frequency band. The good news
 is that the amplitude of the stochastic background, versus $ G \mu$, in the extended domain
 where it is a rather flat rising line, \textit{increases} when $\epsilon$ decreases.
 The bad news is that, because of the low value of the pulsar frequency band
$f_{psr} \sim 0.1 {\rm yr}^{-1}$, the stochastic signal is quite sensitive to the
left cut-off brought by the step function $\Theta(1 - \theta_m[\alpha,f,z] )$ present in
each burst signal Eq.~(\ref{hfz}). This step function corresponds to saying that
the Fourier series representing the GW amplitude emitted by a periodically oscillating
loop only contains mode numbers  $ \vert m \vert \sim (\theta_m(f,z))^{-3} > 1 $.
 As in the case of individual bursts considered above, the main numerical factor
 (when $z \sim 1$, which is the case on the left of the graph plotting $\Omega_g$
 \textit{vs.} $G \mu$) which determines this cut-off is the product $\alpha f t_0$.
When this product gets smaller than 1, the $\Theta$ function  cuts off the
signal. Numerically, one has, for the pulsar frequency band,
$\alpha f_{psr} t_0 \sim 10^9 \alpha \sim 5 \epsilon (G \mu/ 10^{-10})$.
Therefore, when considering, for instance, the type of values  $G \mu \sim 10^{-10} $
 expected from brane inflation models \cite{Tye1,Tye2,KKLMMT,polchinski},
 we see that  values of $\epsilon \lesssim 10^{-1}$ are sufficient for cutting off
 the signal in the pulsar-timing band. On the other hand, for larger values
 of the string tension, the stochastic signal will instead increase when $\epsilon$ decreases.

\begin{figure}
\begin{center}
\hspace{-0.8cm}
\epsfig{file=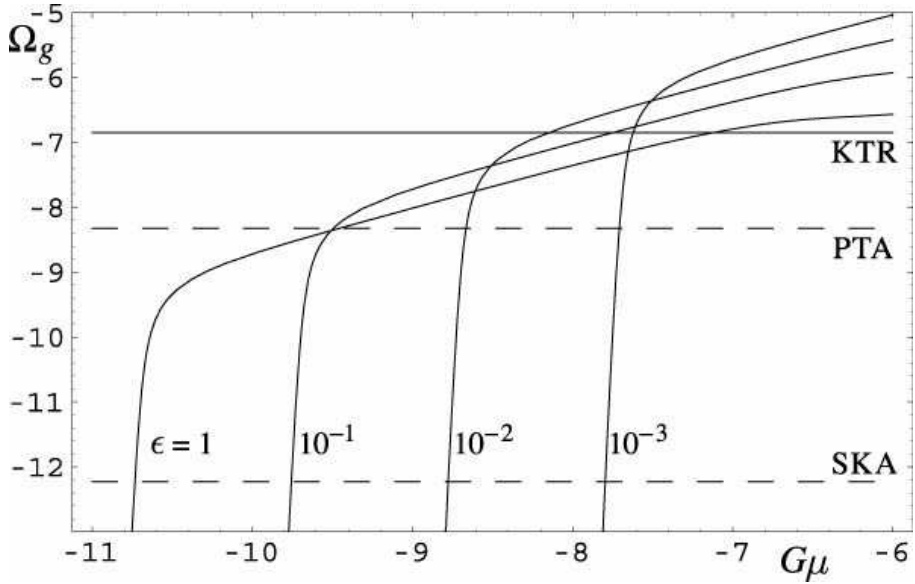,width=12cm}
\medskip
\caption{\sl Effect of a smaller fractional loop-length parameter
$10^{-3} \leq \epsilon \equiv \alpha/(50 G \mu) \leq 1$ on the
fractional contribution $\Omega_g(f_{psr})$ (around the frequency $f_{psr} \sim 1/(10 {\rm yr})$) to the cosmological closure density of the
stochastic GW noise due to overlapping GW bursts emitted by a network of strings.
[Base-$ 10$ log-log plot.] The upper (solid) horizontal line indicates the upper limit
$\Omega_g  < 6  h^{-2}\times 10^{-8}$ derived from 8 years of high-precision
timing of two millisecond pulsars: PSR  1855+09 and PSR 1937+21.
The middle (dashed) horizontal line
indicates the potential sensitivity of  17 years of high-precision timing of PSR  1855+09
(see  text). The lower (dashed) horizontal line indicates the
expected sensitivity from the timing of the set of pulsars to be hopefully detected
by a square-kilometer-array of radio telescopes. }
\label{Fig6}
\end{center}
\end{figure}

In order to control analytically the values of the string tension that could be
detected in pulsar timing experiments,
 it is useful to derive an approximate analytical approximation for the stochastic
signal   $\Omega_g^{confusion}(f) \sim (3 \pi^2/2) (f t_0)^2 h^2_{confusion}(f)$.
By looking at the integrand of Eq.~(\ref{hconfusion}), one can see that
in the low-frequency part of the spectrum which is relevant for the millisecond
pulsar observations (gently rising lines in Figs. 5, 6), this background is produced by the loops radiating at
redshift $z \sim 1$, i.e. within the present Hubble radius, $t\sim t_0$. The value of the integral
Eq.~(\ref{hconfusion}) can  then be approximately estimated\footnote{Note, however,
that this estimate neglects a subdominant, but significant ``floor'' contribution
coming from the radiation era, $z > z_{eq}$} by replacing the integral  $\int dz/z$ by
the value of the integrand at $z \sim1$. This leads to
\beq
h^2_{confusion}(f) \sim \frac{10^2}{\Gamma} c p^{-1} \epsilon^{-1/3} G \mu (\Gamma G \mu)^{-1/3}
(f t_0)^{-7/3},
\eeq
and therefore, using Eq.~(\ref{omega=h2}), to
\beq
\Omega_g(f) \sim 30 G\mu cp^{-1}\epsilon^{-1/3}( \Gamma G \mu f t_0)^{-1/3},
\label{Omega<}
\eeq
where the numerical factor $30$ comes from the combination
$3 \pi^2 10^2/(2 \Gamma)$.
Note that, as was exhibited in Figs. 5, 6,
a decrease in either $p$ or $\epsilon$ from their standard model values ($p=\epsilon=1$) increases
the intensity of the background.
Let us recall that these results are only valid on the gently rising slope on
the left of Figs. 5, 6 , and in particular that they cannot be applied in the
domain $\alpha f t_0 <1$, corresponding to the left cut-off apparent in the figures.
As said above, at frequencies relevant for millisecond pulsar measurements, $f\sim 0.1$ yr$^{-1}$,
this cut off sets in when $\alpha  < 10^{-9}$.

For completeness, let us also sketch a direct derivation of the result (\ref{Omega<}) based on
keeping track of the total  energy emitted by the network of strings.
A loop of length $l$ radiates at a discrete set of frequencies, $f_m=2m/l$, but for large enough $m$
it is well approximated by the continuous spectrum
\beq
dP_g/df\sim \Gamma G\mu^2 l(fl)^{-4/3}.
\label{dp/df}
\eeq
The slow fall-off with frequency $\propto f^{-4/3}$ holds when a cusp forms during a loop oscillation.
For order of magnitude estimates, this formula can be used even for $m\sim 1$, but one has to remember
that the spectrum is cut off at $f \lesssim 1/l$.

Assuming first that $\alpha\ll \Gamma G\mu$, the total number of loops produced in our Hubble volume per Hubble time is
$N\sim 1/(p\alpha)$. Each loop radiates for a time period $\tau\sim (\alpha/\Gamma G\mu)t_0$, so the GW energy
density from all loops (with cusps)  that radiated during the present Hubble time is (per octave of frequency)
\beq
f{d\rho_g\over{df}}\sim f{dP_g\over{df}}\tau {c\over{p\alpha t_0^3}}
\sim {\mu c\over{pt_0^2}}(f\alpha t_0)^{-1/3}.
\eeq
Here, the factor $c$ is added because it measures the fraction of the loops exhibiting cusps.
Expressing this in units of the critical density, $\rho_c\sim 1/(6\pi Gt_0^2)$, we get
for $\Omega_g(f)\equiv( f/\rho_c) d\rho_g/df$ the result
\beq
\Omega_g(f) \sim 6 \pi G\mu cp^{-1}\epsilon^{-1/3}( \Gamma G \mu f t_0)^{-1/3},
\eeq
which is in good agreement (modulo a factor $\sim 0.63$) with the estimate (\ref{Omega<}) above.
Let us recall that all our estimates try to keep the possibly important powers of $2 \pi$, but neglect
various factors ``of order 2''. [We note in this respect that $\Gamma \sim 50$ comes
essentially from a factor $ (2 \pi)^2$, while the factor $10^2$ in Eq.~(\ref{n(fz)}) came from a
factor $\propto 54 \pi$.]

For completeness, let us mention that
the case of $\alpha\gg \Gamma G\mu$ has been reviewed in \cite{book}; the result is
\beq
\Omega_g(f) \sim 6\pi G\mu cp^{-1}(f\alpha_0 t_0)^{-1/3},
\label{Omega>}
\eeq
where $\alpha_0 \equiv \Gamma G \mu$ as above.

Consistently with what we said above, the result (\ref{Omega>}), corresponding to
$\epsilon >1$, can be obtained from the result (\ref{Omega<}), corresponding to
$\epsilon <1$, by replacing $\epsilon \to \epsilon_{eff} =1$ in the latter result.
In order to treat both cases together we can therefore replace everywhere
$\epsilon$ by, say, $\epsilon_{eff} \equiv \epsilon/(1 + \epsilon)$.
With this notation, our approximate analytical result (\ref{Omega<}) numerically yields
\beq
\Omega_g(f) h^2 \sim 10^{-2.46} c(G\mu)^{2/3} p^{-1} \epsilon_{eff}^{-1/3} (f/f_{psr})^{-1/3}.
\label{bound1}
\eeq
Alternatively, any pulsar timing sensitivity level $\Omega_g(f) h^2$ corresponds to a bound
on the string tension of order
\beq
G \mu \sim 10^{3.7} (\Omega_g(f) h^2)^{3/2} (f/f_{psr})^{1/2} c^{-3/2} p^{3/2} \epsilon_{eff}^{1/2}.
\label{bound2}
\eeq
Let us recall that $c \lesssim 1$ denotes the number of cusps occurring per loop oscillation.
The parameter $c$ is expected to be $\sim 1$ for generic smooth
loops \cite{Turok}, but the
presence of many ``kink'' discontinuities along the loop might decrease the (effective) value of $c$
below 1. It seems, however, reasonable to assume that  $c\gtrsim 0.1$.

{}From Eq.~(\ref{bound2}), taking into account that in all cases we have
$p^{3/2} \epsilon_{eff}^{1/2} \lesssim 1$,  we get the inequality
$c^{3/2} G \mu \lesssim 10^{3.7} (\Omega_g(f) h^2)^{3/2} (f/f_{psr})^{1/2} $.
Therefore the firm upper limit $\Omega_{g} h^2 < 6 \times 10^{-8}$ \cite{KTR94} obtained
for a frequency $ f \sim 1/(7 {\rm yr})$, yields the upper limit $c^{3/2} G \mu \lesssim 10^{-7} $,
already quoted in the Introduction. Let us also consider the potential detectability ranges
of pulsar timing arrays. The
  sensitivity level of the current realization of the pulsar timing array (PTA),
$\Omega_g h^2 = 2 \times 10^{-9}$ for $f \sim 1/(15.6 {\rm yr})$ (see above), corresponds to
$G \mu \sim 3.6 \times 10^{-10} c^{-3/2} p^{3/2} \epsilon_{eff}^{1/2}$. This is an impressive
number which shows the vast discovery potential of pulsar timing experiments, and, in particular, the possibility
for PTA experiments to detect the type of string tensions expected from recent superstring
cosmology models \cite{KKLMMT,polchinski}. One must, however, keep in mind
the caveat exhibited in Fig.  6 about the adverse cut-off effect of having $\epsilon < 1$ to which
pulsar experiments are especially sensitive. Fig. 6 shows that more sensitive PTA experiments,
such as the ultimate
Square Kilometer Array PTA, able to probe $\Omega_g h^2 =  10^{-12.6}$, will become
limited to the level
$G \mu \sim 2 \times 10^{-11} \epsilon^{-1}$, by the ability of pulsar timing to probe the frequencies
emitted by string loops.

\section{Conclusions}

It has been argued \cite{Tye1,Tye2,Dvali,KKLMMT,polchinski} that F-
and D-string networks can naturally be formed at the end of brane
inflation \cite{DvaliTye}, with string tensions in the range $10^{-11}
\lesssim G \mu \lesssim 10^{-6}$.
We focussed on models where only one type of string (F or D) is
formed, and considered the gravitational wave (GW) signatures of
cosmological networks of such strings. We studied how the finding
\cite{DV1,DV2} that GW bursts emitted from cusps of oscillating loops
should be detectable by LIGO and LISA interferometers for values of $G
\mu$ as small as $10^{-13}$ might be modified by two separate effects.
First, the reconnection probability $p$ for intersecting F or D
strings might be, contrary to the case of ordinary field-theory
strings, significantly smaller than 1 \cite{Dvali,Tye2,polchinski2}.
Second, Refs.~\cite{DV1,DV2} had assumed that the characteristic size
of newly formed loops was $ l(t) \sim \alpha t$, with $\alpha \sim 50
G \mu$, as expected from from standard GW radiation-reaction arguments
\cite{BB90}. However, recent analyses \cite{Olum1,Olum2} have
suggested that gravitational radiation is less efficient than
originally thought, and might result in a much smaller typical size
for newly formed loops: $ l(t) \sim \alpha t$, where $\alpha \sim
\epsilon 50 G \mu$, with $\epsilon \ll 1$.

A detailed analysis of the effects of the two parameters $p$ and
$\epsilon$ on on the detectability of GW bursts by LIGO or LISA has
shown that the results of \cite{DV1,DV2} are \textit{quite robust}, at
least when $\epsilon \gtrsim 10^{-11}$ in the case of LIGO, or
$\epsilon \gtrsim 10^{-7}$ in the case of LISA. See Figs. 1-4.  In
particular, it is notable that a smaller reconnection probability
can \textit{only increase} the detectability of cosmic
superstrings. However, very small values of $\epsilon$ might lead to
cutting off the burst signals in GW detectors when the string tension
is $G \mu \sim 10^{-10}$ or less.

We have also considered the detectability, via pulsar timing observations, of the stochastic GW background
produced by oscillating string loops. When the loop-length parameter $\alpha$ is large enough,
namely $\alpha = 50 \epsilon G \mu > 10^{-9}$, for the GW frequencies emitted by string loops at redshift $z \sim 1$
to fall within the pulsar sensitivity band $ f \sim 1/( 10 {\rm yr})$, we find that the intensity of the GW background
\textit{increases} $\propto p^{-1} \epsilon^{-1/3}$, when either $p$ or $\epsilon$ gets smaller than 1.
In addition, we find that present pulsar timing experiments have the potential of detecting string tensions as small
as $ G \mu \sim 3.6 \times 10^{-10} c^{-3/2} p^{3/2} \epsilon^{1/2}$, where $c$ denotes the number of cusps per
string oscillation. We urge pulsar observers to reanalyze a recently obtained 17-year combined data set to see whether
the large scatter exhibited by a fraction of the data might be due to a transient GW burst activity of some sort.
We note that future versions of the ``pulsar timing array'' \cite{Lommen,Backer1,Backer2,SKA} might further improve the sensitivity of pulsar observations. The ultimate sensitivity of pulsar timing experiments might then become
limited  (by the ability of pulsar timing to probe the frequencies emitted by string loops) to the level
$G \mu \sim 2 \times 10^{-11} \epsilon^{-1}$. In other words,
if the suggestion of \cite{Olum1,Olum2} is confirmed, and the reduced  loop-length parameter
$\epsilon \equiv \alpha/(50 G \mu)$ turns
out to take very small values $\epsilon \ll 1$ (possibly of the form $\epsilon \sim (50 G \mu)^m$, with $m>0$),
and if $ G \mu$ itself takes small values, $ G \mu \sim 10^{-10}$ \cite{KKLMMT,polchinski}, the GW frequency
band probed by pulsar timing experiments might fall in the domain
 $\alpha = 50 \epsilon G \mu <10^{-9}$ where the GW spectrum from loops is cut off. In such a case, only
 higher-frequency GW experiments, such as LISA or, even better, LIGO, might be able to detect GW
 from string loops.

 Our conclusions show that it is urgent to develop a new generation of string network simulations
 able to determine how the crucial loop-length parameter $\alpha$ depends on $G\mu$ and $p$.
 It would also be quite important to determine the average number $c$ of cusp events per loop oscillation.
 It is only when $\alpha$ and $c$ are known that it will be possible to discuss with any reliability the detectability
 of cosmic superstrings by GW experiments.

\acknowledgements

We wish to thank Michael Kramer for informative discussions about pulsar timing arrays,
and for kindly providing a copy of the second reference in \cite{Lommen}. A. V. thanks IHES
for its hospitality during the inception of this work.

\end{document}